\documentclass{svjour3}                     
\smartqed  
\usepackage{graphicx}
\usepackage[caption=false]{subfig}
\usepackage[numbers]{natbib}
\usepackage{color}
\usepackage{float}
\usepackage{multirow}
\usepackage{makecell}
\usepackage{pbox}
\usepackage[table,x11names]{xcolor}

\usepackage{amsmath,amssymb}

\usepackage{graphics}

\usepackage{adjustbox}

\graphicspath{{figs/}}

\title{Deep Learning in Medical Image Registration: A Survey}

\author{Grant Haskins \and Uwe Kruger \and 	Pingkun~Yan*
\thanks{This work was partially supported by NIH/NIBIB under awards R21EB028001 and R01EB027898, and NIH/NCI under a Bench-to-Bedside award.}%
\thanks{This is a pre-print of an article published in Machine Vision and Applications. The final authenticated version is available online at: https://doi.org/10.1007/s00138-020-01060-x}
}

\institute{G. Haskins, U. Kruger, P. Yan*
	\at
	Department of Biomedical Engineering, Rensselaer Polytechnic Institute, Troy, NY 12180, USA\\
	Asterisk indicates corresponding author\\
	Tel.: +1-518-276-4476\\
	\email{yanp2@rpi.edu}
}

\begin{document}

\newcommand{\absdiv}[1]{%
  \par\addvspace{.5\baselineskip}
  \noindent\textbf{#1}\quad\ignorespaces
}

\maketitle

\begin{abstract}

The establishment of image correspondence through robust image registration is critical to many clinical tasks such as image fusion, organ atlas creation, and tumor growth monitoring, and is a very challenging problem. Since the beginning of the recent deep learning renaissance, the medical imaging research community has developed deep learning based approaches and achieved the state-of-the-art in many applications, including image registration. The rapid adoption of deep learning for image registration applications over the past few years necessitates a comprehensive summary and outlook, which is the main scope of this survey. This requires placing a focus on the different research areas as well as highlighting challenges that practitioners face. This survey, therefore, outlines the evolution of deep learning based medical image registration in the context of both research challenges and relevant innovations in the past few years. Further, this survey highlights future research directions to show how this field may be possibly moved forward to the next level.

\end{abstract}

\section{INTRODUCTION}

Image registration is the process of transforming different image datasets into one coordinate system with matched imaging contents, which has significant applications in medicine. Registration may be necessary when analyzing a pair of images that were acquired from different viewpoints, at different times, or using different sensors/modalities \cite{hill2001medical,zitova2003image}. Until recently, image registration was mostly performed manually by clinicians. However, many registration tasks can be quite challenging and the quality of manual alignments are highly dependent upon the expertise of the user, which can be clinically disadvantageous. To address the potential shortcomings of manual registration, automatic registration has been developed. Although other methods for automatic image registration have been extensively explored prior to (and during) the deep learning renaissance, deep learning has changed the landscape of image registration research \cite{ambinder2005history}. Ever since the success of AlexNet in the ImageNet challenge of 2012 \cite{alom2018history}, deep learning has allowed for state-of-the-art performance in many computer vision tasks including, but not limited to: object detection \cite{ren2015faster}, feature extraction \cite{he2016deep}, segmentation \cite{ronneberger2015u}, image classification \cite{alom2018history}, image denoising \cite{yang2018low}, and image reconstruction \cite{yao2018deep}.

Initially, deep learning was successfully used to augment the performance of iterative, intensity based registration \cite{cheng2018deep, haskins2018learning, simonovsky2016deep}. Soon after this initial application, several groups investigated the intuitive application of reinforcement learning to registration \cite{liao2017artificial, ma2017multimodal, miao2017dilated}. Further, demand for faster registration methods later motivated the development of deep learning based one-step transformation estimation techniques and challenges associated with procuring/generating ground truth data have recently motivated many groups to develop unsupervised frameworks for one-step transformation estimation \cite{de2018deep, li2018non}. One of the hurdles associated with this framework is the familiar challenge of image similarity quantification \cite{heinrich2012mind, viola1997alignment}. Recent efforts that use information theory based similarity metrics \cite{de2018deep}, segmentations of anatomical structures \cite{hu2018weakly}, and generative adversarial network like frameworks \cite{fan2018adversarial} to address this challenge have shown promising results.

\begin{figure}[tbp]
	\centering
	\includegraphics[width=\columnwidth]{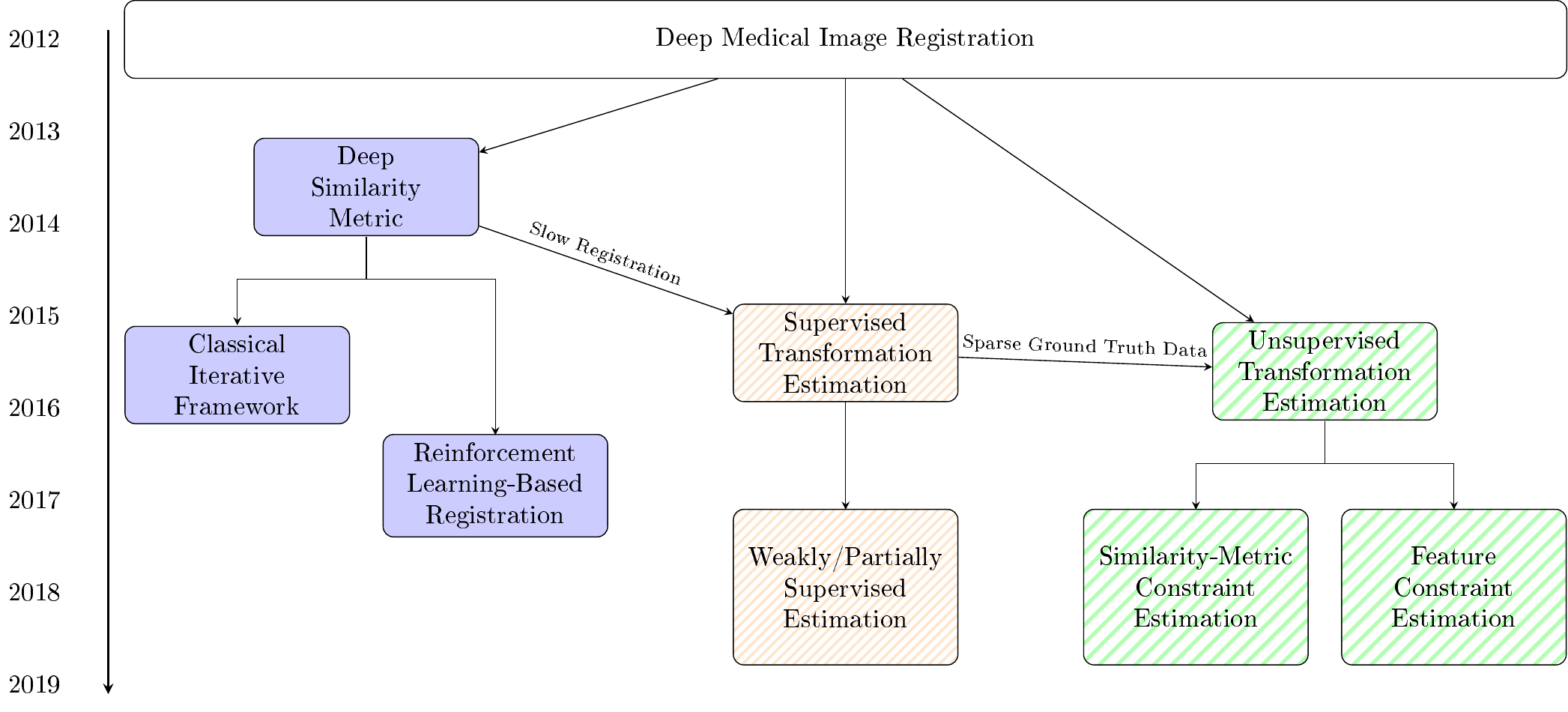}
	\caption{An overview of deep learning based medical image registration broken down by approach type. The popular research directions are written in bold.}
	\label{fig:reg_overview}       
\end{figure}

\begin{figure}[tbp]
	\centering
	\includegraphics[width=\columnwidth]{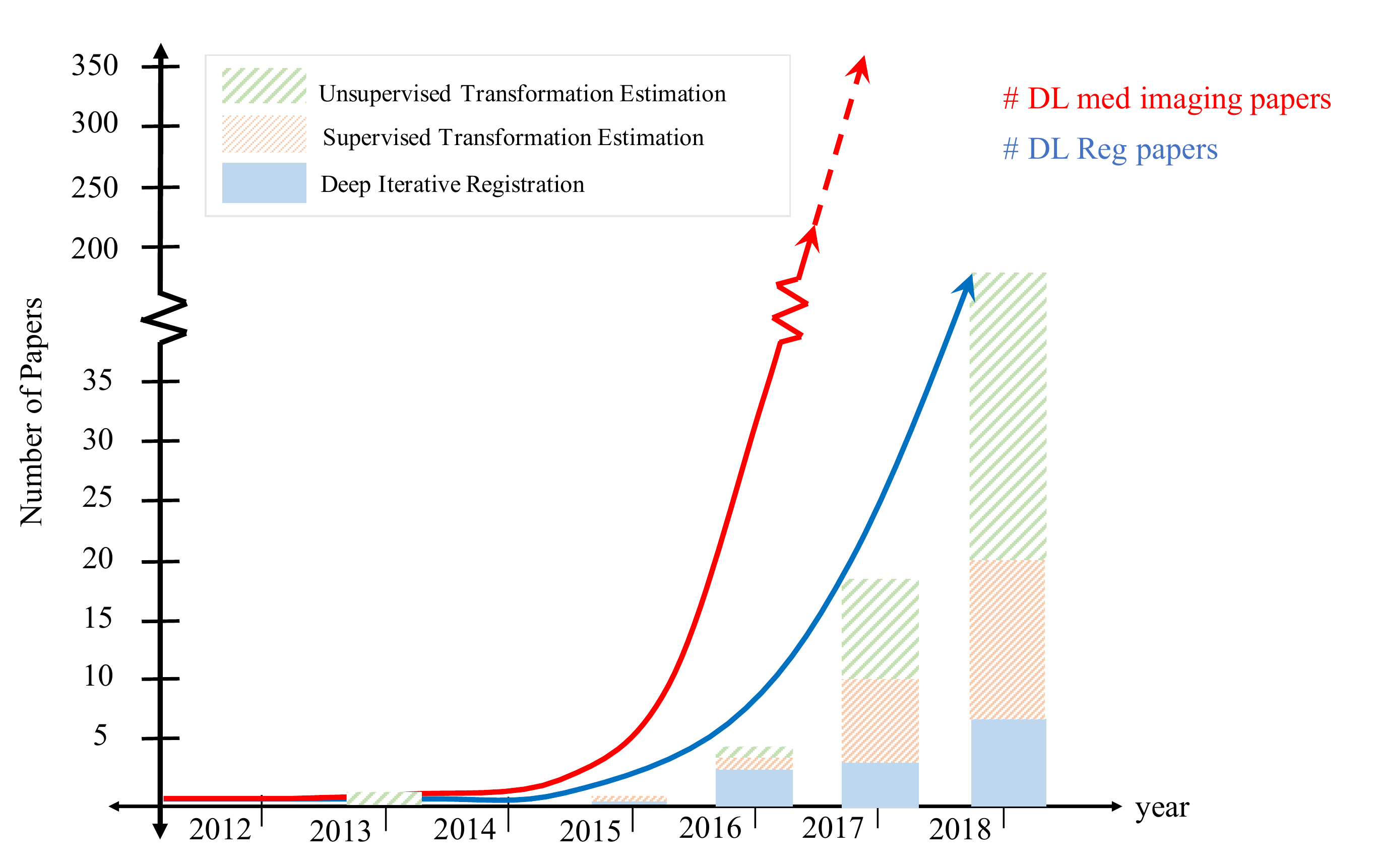}
	\caption{An overview of the number of deep learning based image registration works and deep learning based medical imaging works. The red line represents the trend line for medical imaging based approaches and the blue line represents the trend line for deep learning based medical image registration approaches. The dotted line represents extrapolation.}
\label{fig:overview}       
\end{figure}

Figure \ref{fig:reg_overview} shows the various categorizations of different deep learning based registration methods. On the other hand, Figure \ref{fig:overview} shows the observed growing interest in deep learning based registration methods according to the number of published papers in recent years. As the trends visualized in Figures \ref{fig:reg_overview} and \ref{fig:overview} suggest, this field is moving very quickly to surmount the hurdles associated with deep learning based medical image registration and several groups have already enjoyed significant successes for their applications \cite{ hu2018weakly, liu2018applications,simonovsky2016deep}.

Therefore, the purpose of this article is to comprehensively survey the field of deep learning based medical image registration, highlight common challenges that practitioners face, and discuss future research directions that may address these challenges. Deep learning belongs to a class of machine learning that uses neural networks with a large number of layers to learn representations of data \cite{goodfellow2016deep,schmidhuber2015deep}. When discussing neural networks it is important to provide insight into the different types of neural networks that can be used for various applications, the notable architectures that were recently invented to tackle engineering problems, and the variety of strategies that are used for training neural networks. Therefore, this deep learning introduction section is divided into three sections: Neural Network Types, Network Architectures, and Training Paradigms and Strategies. Note that there are many publicly available libraries that can be used to build the networks described in the section, for example TensorFlow \cite{abadi2016tensorflow}, MXNet \cite{chen2015mxnet}, Keras \cite{chollet2015keras}, Caffe \cite{jia2014caffe}, and PyTorch \cite{paszke2017automatic}.
Detailed discussion of deep learning based medical image analysis and various deep learning research directions is outside of the scope of this article. Comprehensive review articles that survey the application of deep learning to medical image analysis \cite{lee2017deep, litjens2017survey}, reinforcement learning \cite{kaelbling1996reinforcement}, and the application of GANs to medical image analysis \cite{kazeminia2018gans} are recommended to the interested readers. In this article, the surveyed methods were divided into the following three categories: Deep Iterative Registration, Supervised Transformation Estimation, and Unsupervised Transformation Estimation.
Following a discussion of the methods that belong to each of the aforementioned categories, future research directions and current trends are discussed in Section~\ref{sec:trends}.

\section{Deep Iterative Registration}
\label{sec:similarity}

\begin{table*}
    \caption{Deep Iterative Registration Methods Overview. RL denotes reinforcment learning.}
    \label{tab:img_similarity_tab}       
    \centering
    \begin{adjustbox}{}
        \begin{tabular}{|l|l|l|l|l|l|}
            \hline
            \textbf{Ref} & \textbf{Learning} & \textbf{Transform} & \textbf{Modality} & \textbf{ROI} & \textbf{Model} \\
            \hline

            \cite{eppenhof2018error} & Metric & Deformable & CT & Thorax & 9-layer CNN \\
            
            \hline
            
             \cite{blendowski2018combining} & Metric & Deformable & CT & Lung & FCN \\
             \hline
            
            \cite{simonovsky2016deep} & Metric & Deformable & MR & Brain & 5-layer CNN \\
            
            \hline                                                       \cite{wu2013unsupervised} & Metric & Deformable & MR & Brain & 2-layer CAE \\
            
            \hline        
            
            \cite{cheng2018deep} & Metric & Deformable & CT/MR & Head & 5-layer DNN \\  
            
            \hline
            \cite{sedghi2018semi} & Metric & Rigid & MR/US & Abdominal & 5-layer CNN \\
            
            \hline
            \cite{haskins2018learning} & Metric & Rigid & MR/US & Prostate & 14-layer CNN \\
            
            \hline
            \cite{matthewlstm} & Metric & Rigid & MR/US & Fetal Brain & LSTM/STN \\
            \hline
            
            \cite{krebs2017robust} & RL Agent & Deformable & MR & Prostate &  8-layer CNN \\
            
            \hline
            
            \multirow{2}*{\cite{liao2017artificial}} & \multirow{2}*{RL Agent} & \multirow{2}*{Rigid} & \multirow{2}*{CT/CBCT} & Spine/ & \multirow{2}*{8-layer CNN} \\
            &  &  &  & Cardiac & \\
            \hline
            
            \multirow{2}*{\cite{miao2017dilated}} & Multiple & \multirow{2}*{Rigid} & \multirow{2}*{X-ray/CT} & \multirow{2}*{Spine} & \multirow{2}*{Dilated FCN} \\
            & RL Agents &  &  &  & \\
            \hline
            
            \multirow{2}*{\cite{ma2017multimodal}} & \multirow{2}*{RL Agent} & \multirow{2}*{Rigid} & \multirow{2}*{MR/CT} & \multirow{2}*{Spine} & Dueling \\
            &  &  &  &  & Network \\
            \hline
                    
        \end{tabular}
    \end{adjustbox}
\end{table*}

Automatic intensity-based image registration requires both a metric that quantifies the similarity between a moving image and a fixed image and an optimization algorithm that updates the transformation parameters such that the similarity between the images is maximized. Prior to the deep learning renaissance, several manually crafted metrics were frequently used for such registration applications, including: sum of squared differences (SSD), cross-correlation (CC), mutual information (MI) \cite{maes1997multimodality,viola1997alignment}, normalized cross correlation (NCC), and normalized mutual information (NMI). Early applications of deep learning to medical image registration are direct extensions of the intensity-based registration framework \cite{simonovsky2016deep, wu2013unsupervised, wu2016scalable}.
Several groups later used a reinforcement learning paradigm to iteratively estimate a transformation \cite{, krebs2017robust, liao2017artificial, ma2017multimodal, miao2017dilated} because this application is more consistent with how practitioners perform registration.

A description of both types of methods is given in Table~\ref{tab:img_similarity_tab}. We will survey earlier methods that used deep similarity based registration in Section \ref{sec:Intensity} and then some more recently developed methods that use deep reinforcement learning based registration in Section \ref{sec:Deep RL}.

\begin{figure}[tbp]
	\centering
	\includegraphics[width=\columnwidth]{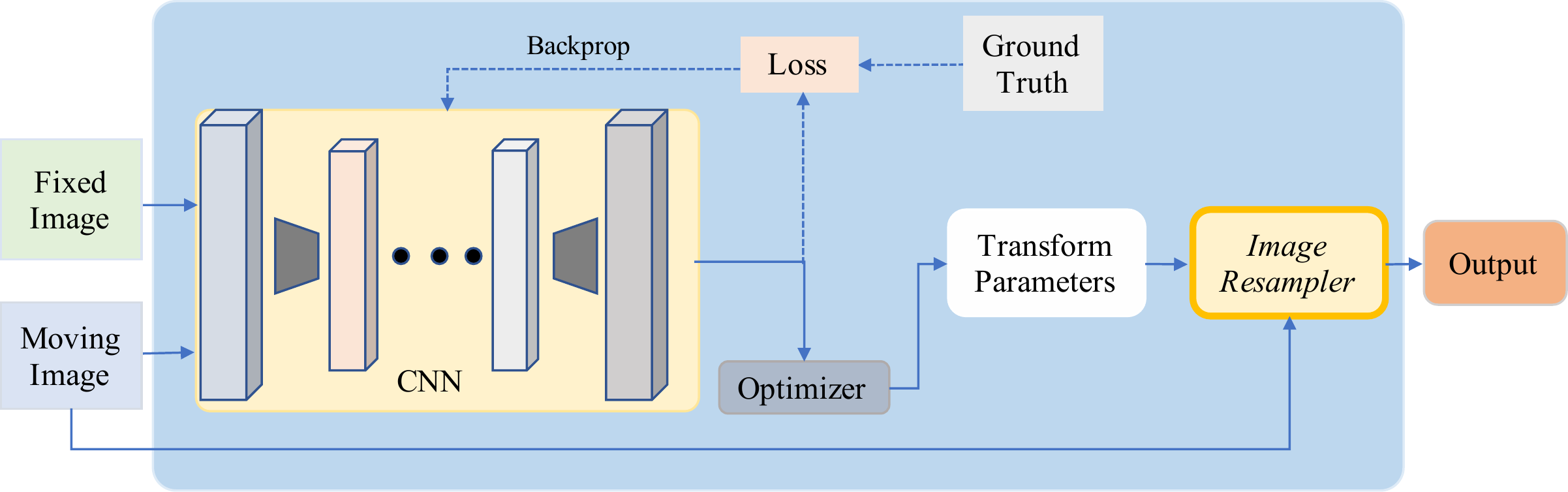}
	\caption{A visualization of the registration pipeline for works that use deep learning to quantify image similarity in an intensity-based registration framework.
}
	\label{fig:img_similarity}       
\end{figure}

\subsection{Deep Similarity based Registration}
\label{sec:Intensity}

In this section, methods that use deep learning to learn a similarity metric are surveyed. This similarity metric is inserted into a classical intensity-based registration framework with a defined interpolation strategy, transformation model, and optimization algorithm. A visualization of this overall framework is given in Fig.~\ref{fig:img_similarity}. The solid lines represent data flows that are required during training and testing, while the dashed lines represent data flows that are required only during training. Note that this is the case for the remainder of the figures in this article as well.

\subsubsection{Overview of Works}
\label{sec:Overview Intensity}

Although manually crafted similarity metrics perform reasonably well in the unimodal registration case, deep learning has been used to learn superior metrics. This section will first discuss approaches that use deep learning to augment the performance of unimodal intensity based registration pipelines before multimodal registration.  


\paragraph{Unimodal Registration}
Wu et al. \cite{wu2013unsupervised, wu2016scalable} were the first to use deep learning to obtain an application specific similarity metric for registration. They extracted the features that are used for unimodal, deformable registration of 3D brain MR volumes using a convolutional stacked autoencoder (CAE). They subsequently performed the registration using gradient descent to optimize the NCC of the two sets of features. This method outperformed diffeomorphic demons \cite{vercauteren2009diffeomorphic} and HAMMER \cite{shen2007image} based registration techniques.


Recently, Eppenhof et al. \cite{eppenhof2018error} estimated registration error for the deformable registration of 3D thoracic CT scans (inhale-exhale) in an end-to-end capacity. They used a 3D CNN to estimate the error map for inputted inhale-exhale pairs of thoracic CT scans. Like the above method, only learned features were used in this work.

Instead, Blendowski et al. \cite{blendowski2018combining} proposed the combined use of both CNN-based descriptors and manually crafted MRF-based self-similarity descriptors for lung CT registration. Although the manually crafted descriptors outperformed the CNN-based descriptors, optimal performance was achieved using both sets of descriptors. This indicates that, in the unimodal registration case, deep learning may not outperform manually crafted methods. However, it can be used to obtain complementary information.

\paragraph{Multimodal Registration}
The advantages of the application of deep learning to intensity based registration are more obvious in the multimodal case, where manually crafted similarity metrics have had very little success. 


Cheng et al. \cite{chengdeep, cheng2018deep} recently used a stacked denoising autoencoder to learn a similarity metric that assesses the quality of the rigid alignment of CT and MR images. They showed that their metric outperformed NMI-optimization-based and local cross correlation (LCC)-optimization-based for their application.

In an effort to explicitly estimate image similarity in the multimodal case, Simonovsky et al. \cite{simonovsky2016deep} used a CNN to learn the dissimilarity between aligned 3D T1 and T2 weighted brain MR volumes. Given this similarity metric, gradient descent was used in order to iteratively update the parameters that define a deformation field. This method was able to outperform MI-optimization-based registration and set the stage for deep intensity based multimodal registration.

Additionally, Sedghi et al. \cite{sedghi2018semi} performed the rigid registration of 3D US/MR (modalities with an even greater appearance difference than MR/CT) abdominal scans by using a 5-layer neural network to learn a similarity metric that is then optimized by Powell’s method. 
This approach also outperformed MI-optimization-based registration. Haskins et al. \cite{haskins2018learning} learned a similarity metric for multimodal rigid registration of MR and transrectal US (TRUS) volumes by using a CNN to predict target registration error (TRE). Instead of using a traditional optimizer like the above methods, they used an evolutionary algorithm to explore the solution space prior to using a traditional optimization algorithm because of the learned metric's lack of convexity. This registration framework outperformed MIND-optimization-based \cite{heinrich2012mind} and MI-optimization-based registration.
In stark contrast to the above methods, Wright et al. \cite{matthewlstm} used LSTM spatial co-transformer networks to iteratively register MR and US volumes group-wise. The recurrent spatial co-transformation occurred in three steps: image warping, residual parameter prediction, parameter composition. They demonstrated that their method is more capable of quantifying image similarity than a previous multimodal image similarity quantification method that uses self-similarity context descriptors \cite{heinrich2013towards}.

\subsubsection{Discussion and Assessment}
\label{sec:Discussion Intensity}

Recent works have confirmed the ability of neural networks to assess image similarity in multimodal medical image registration. The results achieved by the approaches described in this section demonstrate that deep learning can be successfully applied to challenging registration tasks. However, the findings from \cite{blendowski2018combining} suggest that learned image similarity metrics may be best suited to complement existing similarity metrics in the unimodal case. Further, it is difficult to use these iterative techniques for real time registration. 

\subsection{Reinforcement Learning based Registration}
\label{sec:Deep RL}

\begin{figure}[tbp]
	\centering
	\includegraphics[width=\columnwidth]{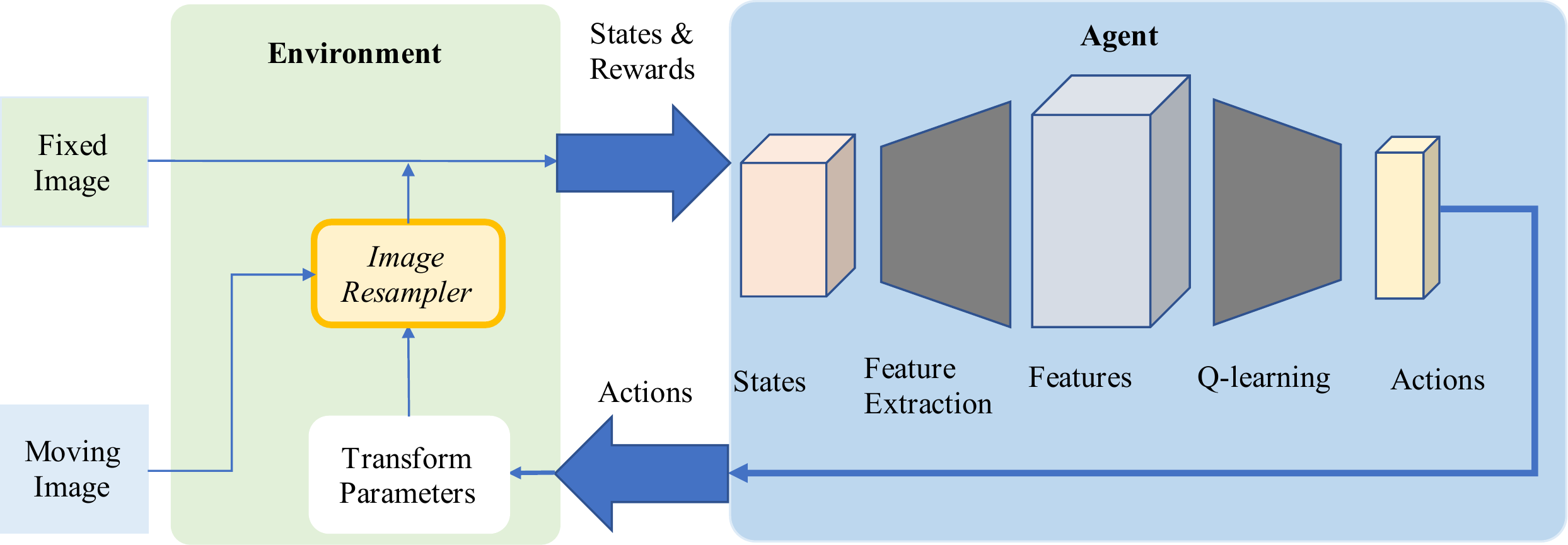}
	\caption{A visualization of the registration pipeline for works that use deep reinforcement learning to implicitly quantify image similarity for image registration. Here, an agent learns to map states to actions based on rewards that it receives from the environment.
}
	\label{fig:RL_reg}       
\end{figure}

In this section, methods that use reinforcement learning for their registration applications are surveyed. Here, a trained agent is used to perform the registration as opposed to a pre-defined optimization algorithm. A visualization of this framework is given in Fig.~\ref{fig:RL_reg}. Reinforcement learning based registration typically involves a rigid transformation model. However, it is possible to use a deformable transformation model. 

Liao et al. \cite{liao2017artificial} were the first to use reinforcment learning based registration to perform the rigid registration of cardiac and abdominal 3D CT images and cone-beam CT (CBCT) images. They used a greedy supervised approach for end-to-end training with an attention-driven hierarchical strategy. Their method outperformed MI based registration and semantic registration using probability maps.

Shortly after, Kai et al. \cite{ma2017multimodal} used a reinforcement learning approach to perform the rigid registration of MR/CT chest volumes. This approach is derived from $Q$-learning and leverages contextual information to determine the depth of the projected images. The network used in this method is derived from the dueling network architecture \cite{wang2015dueling}. Notably, this work also differentiates between terminal and non-terminal rewards. This method outperforms registration methods that are based on iterative closest points (ICP), landmarks, Hausdorff distance, Deep Q Networks, and the Dueling Network \cite{wang2015dueling}.

Instead of training a single agent like the above methods, Miao et al. \cite{miao2017dilated} used a multi-agent system in a reinforcement learning paradigm to rigidly register X-Ray and CT images of the spine. They used an auto-attention mechanism to observe multiple regions and demonstrate the efficacy of a multi-agent system. They were able to significantly outperform registration approaches that used a state-of-the-art similarity metric given by \cite{de20163d}. 

As opposed to the above rigid registration based works, Krebs et al. \cite{krebs2017robust} used a reinforcement learning based approach to perform the deformable registration of 2D and 3D prostate MR volumes. They used a low resolution deformation model for the registration and fuzzy action control to influence the stochastic action selection. The low resolution deformation model is necessary to restrict the dimensionality of the action space. This approach outperformed registration performed using the Elastix toolbox \cite{klein2010elastix} and LCC-Demons \cite{lorenzi2013lcc} based registration techniques.

\begin{figure}[tbp]
	\centering
	\includegraphics[width=\columnwidth]{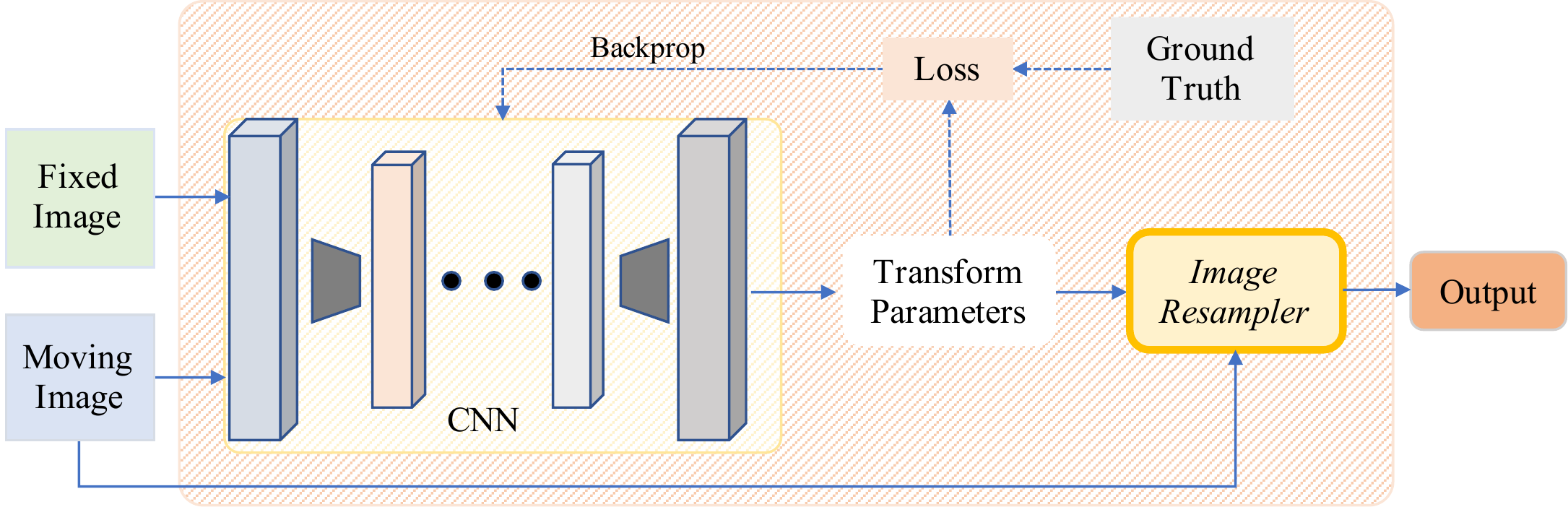}
	\caption{A visualization of supervised single step registration.
}
	\label{fig:trans_est_sup}       
\end{figure}

The use of reinforcement learning is intuitive for medical image registration applications. One of the principle challenges for reinforcement learning based registration is the ability to handle high resolution deformation fields. There are no such challenges for rigid registration. Because of the intuitive nature and recency of these methods, we expect that such approaches will receive more attention from the research community in the next few years.

\begin{table*}
    \caption{Supervised Transformation Estimation Methods. Gray rows use Diffeomorphisms.}
    \label{tab:trans_supervised}       
    \centering
    \begin{adjustbox}{}
        \begin{tabular}{|l|l|l|l|l|l|l|}
            \hline
            \textbf{Ref} & \textbf{Supervision} & \textbf{Transform} & \textbf{Modality} & \textbf{ROI} & \textbf{Model} \\
            \hline
            
            \rowcolor{lightgray} \cite{yang2016fast} & Real Transforms & Deformable & MR & Brain & FCN \\
            \hline
            
            \cite{cao2017deformable} & Real Transforms & Deformable & MR & Brain & 9-layer CNN \\
            \hline
            
            \cite{lv2018respiratory} & Real Transforms & Deformable & MR & Abdominal & CNN \\
            \hline
            
            \cite{rohe2017svf} & Real Transforms & Deformable & MR & Cardiac & SVF-Net \\                
            \hline
            
            \multirow{2}*{\cite{sokooti2017nonrigid}} & Synthetic & \multirow{2}*{Deformable} & \multirow{2}*{CT} & \multirow{2}*{Chest} & \multirow{2}*{RegNet} \\
            
            & Transforms &  &  &  &  \\
            \hline
            
            \multirow{2}*{\cite{eppenhof2018pulmonary}} & Synthetic & \multirow{2}*{Deformable} & \multirow{2}*{CT} & \multirow{2}*{Lung} & \multirow{2}*{U-Net} \\
            
            & Transforms &  &  &  &  \\
            \hline
            
            \multirow{2}*{\cite{uzunova2017training}} & Synthetic & \multirow{2}*{Deformable} & \multirow{2}*{MR} & Brain/ & \multirow{2}*{FlowNet} \\
            
            & Transforms &  &  & Cardiac &  \\
            \hline
            
            \multirow{2}*{\cite{itoautomated}} & Synthetic & \multirow{2}*{Deformable} & \multirow{2}*{MR} & \multirow{2}*{Brain} & \multirow{2}*{GoogleNet} \\
            
            & Transforms &  &  &  &  \\
            \hline
            
            \multirow{2}*{\cite{sun2018towards}} & Synthetic & \multirow{2}*{Deformable} & \multirow{2}*{CT/US} & \multirow{2}*{Liver} & \multirow{2}*{DVFNet} \\
            
            & Transforms &  &  &  &  \\
            \hline
            
            \multirow{2}*{\cite{yang2017uncertainty}} & Real + Synthetic & \multirow{2}*{Deformable} & \multirow{2}*{MR} & \multirow{2}*{Brain} & \multirow{2}*{FCN} \\
            
            & Transforms &  &  &  &  \\
            \hline
            
            \multirow{2}*{\cite{sloan2018learning}} & Synthetic & \multirow{2}*{Rigid} & \multirow{2}*{MR} & \multirow{2}*{Brain} & 6-layer CNN \\
            
            & Transforms &  &  &   & 10-layer FCN \\
            \hline
            
            \multirow{2}*{\cite{salehi2018real}} & Synthetic & \multirow{2}*{Rigid} & \multirow{2}*{MR} & \multirow{2}*{Brain} & 11-layer CNN  \\
            
            & Transforms &  &  &  & ResNet-18 \\
            \hline
            
            \multirow{2}*{\cite{zheng2018pairwise}} & Synthetic & \multirow{2}*{Rigid} & \multirow{2}*{X-ray} & \multirow{2}*{Bone} & 17-layer CNN \\
            
            & Transforms &  &  &  & PDA Module \\
            \hline           
            
            \multirow{2}*{\cite{miao2016real}} & Synthetic & \multirow{2}*{Rigid} & X-ray/ & \multirow{2}*{Bone} & \multirow{2}*{6-layer CNN} \\
            
            & Transforms &  & DDR &  &  \\
            \hline
            
            \multirow{2}*{\cite{chee2018airnet}} & Synthetic & \multirow{2}*{Rigid} & \multirow{2}*{MR} & \multirow{2}*{Brain} & \multirow{2}*{AIRNet} \\
            
            & Transforms &  &  &  &  \\
            \hline
            
            \cite{hu2018weakly} & Segmentations & Deformable & MR/US & Prostate & 30-layer FCN \\
            \hline
            
            \multirow{2}*{\cite{hering2018enhancing}} & Segmentations +& \multirow{2}*{Deformable} & \multirow{2}*{MR/US} & \multirow{2}*{Prostate} & U-Net \\
            
            & Similarity Metric &  &  &  & GAN \\
            \hline
            
            \multirow{2}*{\cite{hu2018adversarial}} & Segmentations +& \multirow{2}*{Deformable} & \multirow{2}*{MR/US} & \multirow{2}*{Prostate} & \multirow{2}*{GAN} \\
            
            & Adversarial Loss &  &  &  & \\

            \hline
            
            \multirow{2}*{\cite{fan2018birnet}} & Real Transforms +& \multirow{2}*{Deformable} & \multirow{2}*{MR} & \multirow{2}*{Brain} & \multirow{2}*{U-Net} \\
            
            & Similarity Metric &  &  &  & \\

            \hline
            
            \multirow{3}*{\cite{yan2018adversarial}} & Synthetic & \multirow{3}*{Rigid} & \multirow{3}*{MR/US} & \multirow{3}*{Prostate} & \multirow{3}*{GAN} \\
            
            & Transforms +&  &  &  & \\
            
            & Adversarial Loss &  &  &  & \\

            \hline
        \end{tabular}
    \end{adjustbox}
\end{table*}

\section{Supervised Transformation Estimation}

Despite the early success of the previously described approaches, the transformation estimation in these methods is iterative, which can lead to slow registration. \cite{haskins2018learning}. This is especially true in the deformable registration case where the solution space is high dimensional \cite{lee2017deep}. This motivated the development of networks that could estimate the transformation that corresponds to optimal similarity in one step. However, fully supervised transformation estimation (the exclusive use of ground truth data to define the loss function) has several challenges that are highlighted in this section. 

A visualization of supervised transformation estimation is given in Fig.~\ref{fig:trans_est_sup} and a description of notable works is given in Table~\ref{tab:trans_supervised}. 
This section first discusses methods that use fully supervised approaches in Section \ref{sec:full} and then discusses methods that use dual/weakly supervised approaches in Section \ref{sec:part}.


\subsection{Fully Supervised Transformation Estimation}
\label{sec:full}

In this section, methods that used full supervision for single-step registration are surveyed. Using a neural network to perform registration as opposed to an iterative optimizer significantly speeds up the registration process. 

\subsubsection{Overview of works}
\label{sec:full overview}

Several registration application require deformable transformation models that often prohibit the use of traditional convolutional neural networks because of the computational expense associated with using FC-layers to make predictions in highly dimensional solution spaces \cite{krebs2017robust}. Because the networks that are used to predict deformation fields are fully convolutional, the dimensionality of the solution space associated with a deformation field does not introduce additional computational constraints \cite{yang2016fast}. This section will first discuss approaches that use a rigid transformation model and then discuss approaches that use a deformable transformation model.

\paragraph{Rigid Registration}
Miao et al. \cite{miao2016cnn, miao2016real} were the first to use deep learning to predict rigid transformation parameters. They used a CNN to predict the transformation matrix associated with the rigid registration of 2D/3D X-ray attenuation maps and 2D X-ray images. Hierarchical regression is proposed in which the 6 transformation parameters are partitioned into 3 groups. Ground truth data was synthesized in this approach by transforming aligned data. This is the case for the next three approaches that are described as well. This approach outperformed MI, CC, and gradient correlation (GC)-optimization-based registration approaches with respect to both accuracy and computational efficiency. The improved computational efficiency is due to the use of a forward pass through a neural network instead of an optimization algorithm to perform the registration.

Recently, Chee et al. \cite{chee2018airnet} used a CNN to predict the transformation parameters used to rigidly register 3D brain MR volumes. In their framework, affine image registration network (AIRNet), the MSE between the predicted and ground truth affine transforms is used to train the network. They were able to outperform MI-optimization-based registration for both the unimodal and multimodal cases.

That same year, Salehi et al. \cite{salehi2018real} used a deep residual regression network, a correction network, and a bivariant geodesic distance based loss function to rigidly register T1 and T2 weighted 3D fetal brain MRs for atlas construction. The use of the residual network to initially register the image volumes prior to the forward pass through the correction network allowed for an enhancement of the capture range of the registration. This approach was evaluated for both slice-to-volume registration and volume-to-volume registration. They validated the efficacy of their geodesic loss term and outperformed NCC-optimization-based registration.

Additionally, Zheng et al. \cite{zheng2018pairwise} proposed the integration of a pairwise domain adaptation module (PDA) into a pre-trained CNN that performs the rigid registration of pre-operative 3D X-Ray images and intraoperative 2D X-ray images using a limited amount of training data. Domain adaptation was used to address the discrepancy between synthetic data that was used to train the deep model and real data. 

Sloan et al. \cite{sloan2018learning} used a CNN is used to regress the rigid transformation parameters for the registration of T1 and T2 weighted brain MRs. Both unimodal and multimodal registration were investigated in this work. The parameters that constitute the convolutional layers that were used to extract low-level features in each image were only shared in the unimodal case. In the multimodal case, these parameters were learned separately. This approach also outperformed MI-optimization-based image registration.

\paragraph{Deformable Registration}
Unike the previous section, methods that use both real and synthesized ground truth labels will be discussed. Methods that use clinical/publicly available ground truth labels for training are discussed first. This ordering is reflective of the fact that simulating realistic deformable transformations is more difficult than simulating realistic rigid transformations. 

First, Yang et al. \cite{yang2016fast} predicted the deformation field with an FCN that is used to register 2D/3D intersubject brain MR volumes in a single step. A U-net like architecture \cite{ronneberger2015u} was used in this approach. Further, they used large diffeomorphic metric mapping to provide a basis, used the initial momentum values of the pixels of the image volumes as the network input, and evolved these values to obtain the predicted deformation field. This method outperformed semi-coupled dictionary learning based registration \cite{cao2015semi}.

The following year, Rohe et al. \cite{rohe2017svf} also used a U-net \cite{ronneberger2015u} inspired network to estimate the deformation field used to register 3D cardiac MR volumes. Mesh segmentations are used to compute the reference transformation for a given image pair and SSD between the prediction and ground truth is used as the loss function. This method outperformed LCC Demons based registration \cite{lorenzi2013lcc}. 

That same year, Cao et al. \cite{cao2017deformable} used a CNN to map input image patches of a pair of 3D brain MR volumes to their respective displacement vector. The totality of these displacement vectors for a given image constitutes the deformation field that is used to perform the registration. Additionally, they used the similarity between inputted image patches to guide the learning process. Further, they used equalized active-points guided sampling strategy that makes it so that patches with higher gradient magnitudes and displacement values are more likely to be sampled for training. This method outperforms SyN \cite{avants2008symmetric} and Demons \cite{vercauteren2009diffeomorphic} based registration methods.

Recently, Jun et al. \cite{lv2018respiratory} used a CNN to perform the deformable registration of abdominal MR images to compensate for the deformation that is caused by respiration. This approach achieved registration results that are superior to those obtained using non-motion corrected registrations and local affine registration. Recently, unlike many of the other approaches discussed in this paper, Yang et al. \cite{yang2017uncertainty} quantified the uncertainty associated with the deformable registration of 3D T1 and T2 weighted brain MRs using a low-rank Hessian approximation of the variational gaussian distribution of the transformation parameters. This method was evaulated on both real and synthetic data. 

Just as deep learning practitioners use random transformations to enhance the diversity of their dataset, Sokooti et al. \cite{sokooti2017nonrigid} used random DVFs to augment their dataset. They used a multi-scale CNN to predict a deformation field. This deformation is used to perform intra-subject registration of 3D chest CT images. 
This method used late fusion as opposed to early fusion, in which the patches are concatenated and used as the input to the network. The performance of their method is competitive with B-Spline based registration \cite{sokooti2017nonrigid}.

Such approaches have notable, but also limited ability to enhance the size and diversity of datasets. These limitations motivated the development of more sophisticated ground truth generation. The rest of the approaches described in this section use simulated ground truth data for their applications.

For example, Eppenhof et al. \cite{eppenhof2018pulmonary} used a 3D CNN to perform the deformable registration of inhale-exhale 3D lung CT image volumes. A series of multi-scale, random transformations of aligned image pairs eliminate the need for manually annotated ground truth data while also maintaining realistic image appearance. Further, as is the case with other methods that generate ground truth data, the CNN can be trained using relatively few medical images in a supervised capacity.


Unlike the above works, Uzunova et al. \cite{uzunova2017training} generated ground truth data using statistical appearance models (SAMs). They used a CNN to estimate the deformation field for the registration of 2D brain MRs and 2D cardiac MRs, and adapt FlowNet \cite{dosovitskiy2015flownet} for their application. They demonstrated that training FlowNet using SAM generated ground truth data resulted in superior performance to CNNs trained using either randomly generated ground truth data or ground truth data obtained using the registration method described in \cite{ehrhardt2015variational}. 

Unlike the other methods in this section that use random transformations or manually crafted methods to generate ground truth data, Ito et al. \cite{itoautomated} used a CNN to learn plausible deformations for ground truth data generation. They evaluated their approach on the 3D brain MR volumes in the ADNI dataset and outperformed the MI-optimization-based approach proposed in \cite{ikeda2014efficient}.

\subsubsection{Discussion and Assessment}
\label{sec:full discuss}

Supervised transformation estimation has allowed for real time, robust registration across applications. However, such works are not without their limitations. Firstly, the quality of the registrations using this framework is dependent on the quality of the ground truth registrations. The quality of these labels is, of course, dependent upon the expertise of the practitioner. Furthermore, these labels are fairly difficult to obtain because there are relatively few individuals with the expertise necessary to perform such registrations. Transformations of training data and the generation of synthetic ground truth data can address such limitations. However, it is important to ensure that simulated data is sufficiently similar to clinical data. These challenges motivated the development of partially supervised/unsupervised approaches, which will be discussed next.

\subsection{Dual/Weakly Supervised Transformation Estimation}
\label{sec:part}

Dual supervision refers to the use of both ground truth data and some metric that quantifies image similarity to train a model. On the other hand, weak supervision refers to using the overlap of segmentations of corresponding anatomical structures to design the loss function. This section will discuss the contributions of such works in Section \ref{sec:part overview} and then discuss the overall state of this research direction in Section \ref{sec:part discuss}.

\subsubsection{Overview of works}
\label{sec:part overview}

\begin{figure}[tbp]
	\centering
	\includegraphics[width=\columnwidth]{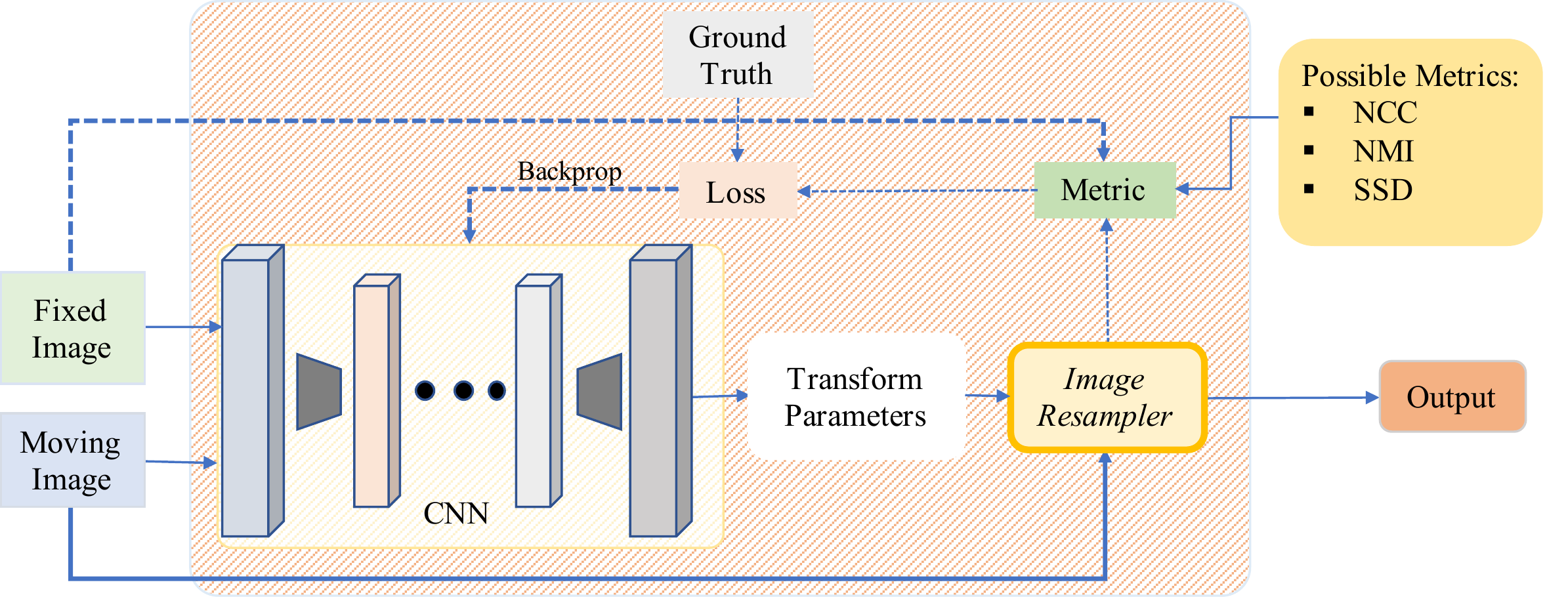}
	\caption{A visualization of deep single step registration where the agent is trained using dual supervision. The loss function is determined using both a metric that quantifies image similarity and ground truth data.}
	\label{fig:trans_est_part}       
\end{figure}

First, this section will discuss methods that use dual supervised and then will discuss methods that use weak supervision. Recently, Fan et al. \cite{fan2018birnet} used hierarchical, dual-supervised learning to predicted the deformation field for 3D brain MR registration. They amend the traditional U-Net architecture \cite{ronneberger2015u} by using ``gap-filling'' (\emph{i.e.}, inserting convolutional layers after the U-type ends or the architecture) and coarse-to-fine guidance. This approach leveraged both the similarity between the predicted and ground truth transformations, and the similarity between the warped and fixed images to train the network. The architecture detailed in this method outperformed the traditional U-Net architecture and the dual supervision strategy is verified by ablating the image similarity loss function term. A visualization of dual supervised transformation estimation is given in Fig.~\ref{fig:trans_est_part}. 

\begin{figure}[tbp]
	\centering
	\includegraphics[width=\columnwidth]{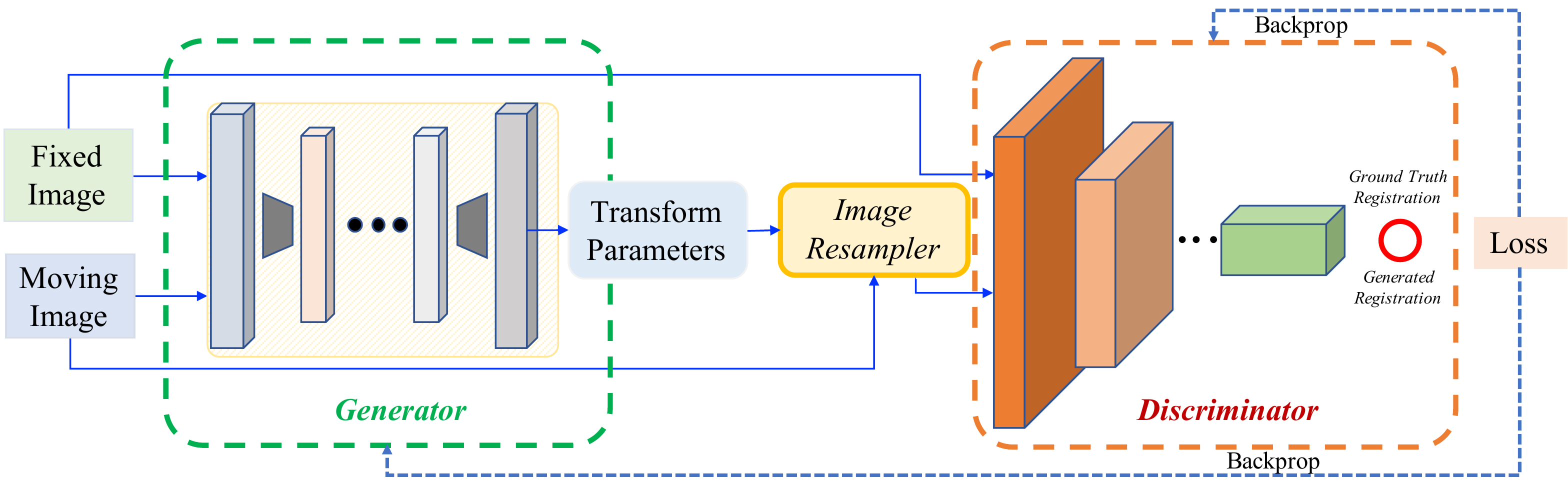}
	\caption{A visualization of an adversarial image registration framework. Here, the generator is trained using output from the discriminator. The discriminator takes the form of a learned metric here.}
	\label{fig:trans_est_adv}       
\end{figure}

On the other hand, Yan et al. \cite{yan2018adversarial} used a framework that is inspired by the GAN \cite{goodfellow2014generative} to perform the rigid registration of 3D MR and TRUS volumes. In this work, the generator was trained to estimate a rigid transformation. While, the discriminator was trained to discern between images that were aligned using the ground truth transformations and images that were aligned using the predicted transformations. Both Euclidean distance to ground truth and an adversarial loss term are used to construct the loss function in this method. Note that the adversarial supervision strategy that was used in this approach is similar to the ones that are used in a number of unsupervised works that will be described in the next section. A visualization of adversarial transformation estimation is given in Fig.~\ref{fig:trans_est_adv}.

\begin{figure}[tbp]
	\centering
	\includegraphics[width=\columnwidth]{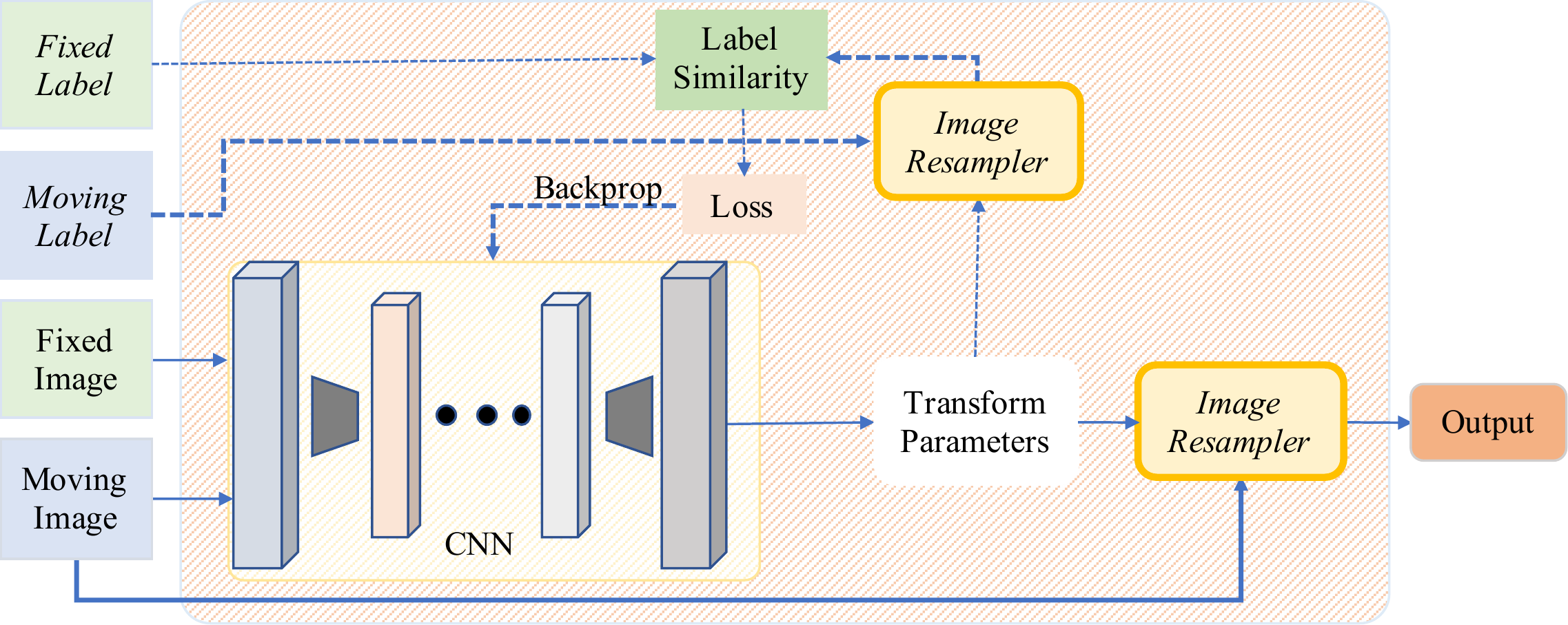}
	\caption{A visualization of deep single step registration where the agent is trained using label similarity (\emph{i.e.} weak supervision). Manually annotated data (segmentations) are used to define the loss function used to train the network.}
	\label{fig:trans_est_weak}       
\end{figure}

Unlike the above methods that used dual supervision, Hu et al. \cite{hu2018label, hu2018weakly} recently used label similarity to train their network to perform MR-TRUS registration. 
In their initial work, they used two neural networks: local-net and global-net to estimate the global affine transformation with 12 degrees of freedom and the local dense deformation field respectively \cite{hu2018label}. The local-net uses the concatenation of the transformation of the moving image given by the global-net and the fixed image as its input. However, in their later work \cite{hu2018weakly}, they combine these networks in an end-to-end framework. This method outperformed NMI-optimization-based and NCC based registration. A visualization of weakly supervised transformation estimation is given in Fig.~\ref{fig:trans_est_weak}. In another work, Hu et al. \cite{hu2018adversarial} simultaneously maximized label similarity and minimized an adversarial loss term to predict the deformation for MR-TRUS registration. This regularization term forces the predicted transformation to result in the generation of a realistic image. Using the adversarial loss as a regularization term is likely to successfully force the transformation to be realistic given proper hyper parameter selection. The performance of this registration framework was inferior to the performance of their previous registration framework described above. However, they showed that adversarial regularization is superior to standard bending energy based regularization.
Similar to the above method, Hering et al. \cite{hering2018enhancing} built upon the progress made with respect to both dual and weak supervision by introducing a label and similarity metric based loss function for cardiac motion tracking via the deformable registration of 2D cine-MR images. Both segmentation overlap and edge based normalized gradient fields distance were used to construct the loss function in this approach. Their method outperformed a multilevel registration approach similar to the one proposed in \cite{ruhaak2013highly}.

\subsubsection{Discussion and Assessment}
\label{sec:part discuss}

Direct transformation estimation marked a major breakthrough for deep learning based image registration. With full supervision, promising results have been obtained. However, at the same time, those techniques require a large amount of detailed annotated images for training. Partially/weakly supervised transformation estimation methods alleviated the limitations associated with the trustworthiness and expense of ground truth labels. However, they still require manually annotated data (\emph{e.g.} ground truth and/or segmentations). On the other hand, weak supervision allows for similarity quantification in the multimodal case. Further, partial supervision allows for the aggregation of methods that can be used to assess the quality of a predicted registration. As a result, there is growing interest in these research areas.

\begin{table*}
	\caption{Unsupervised Transformation Estimation Methods. Grays rows use Diffeomorphisms.}
	\label{tab:trans_unsupervised}       
	\centering
	\begin{adjustbox}{}
		\begin{tabular}{|l|l|l|l|l|l|}
			\hline
			\textbf{Ref} & \textbf{Loss Function} & \textbf{Transform} & \textbf{Modality} & \textbf{ROI} & \textbf{Model} \\
			\hline
            
			\multirow{2}*{\cite{jiang2018cnn}} & \multirow{2}*{SSD} & \multirow{2}*{Deformable} & \multirow{2}*{CT} & \multirow{2}*{Chest} & Multi-scale \\
			&  &  &  &  & CNN \\ \hline
			
			\multirow{2}*{\cite{ghosal2017deep}} & \multirow{2}*{UB SSD} & \multirow{2}*{Deformable} & \multirow{2}*{MR} & \multirow{2}*{Brain} & 19-layer \\
			&  &  &  &  & FCN \\ \hline
			
            \rowcolor{lightgray} \cite{zhang2018inverse} & MSD & Deformable & MR & Brain & ICNet \\ \hline
			
    		\multirow{2}*{\cite{shu2018unsupervised}} & \multirow{2}*{MSE} & \multirow{2}*{Deformable} & \multirow{2}*{SEM} & \multirow{2}*{Neurons} & 11-layer \\
			&  &  &  &  & CNN \\ \hline
			            
            \rowcolor{lightgray} \cite{dalca2018unsupervised} & MSE & Deformable & MR & Brain & VoxelMorph \\ 
			
			\hline
			
			\multirow{2}*{\cite{sheikhjafari2018unsupervised}} & \multirow{2}*{MSE} & \multirow{2}*{Deformable} & \multirow{2}*{MR} & Cardiac & 8-layer \\
			&  &  &  & Cine & FCNet \\ \hline
			
            \cite{kuang2018faim} & CC & Deformable & MR & Brain & FAIM \\
			\hline 
			
			\multirow{2}*{\cite{li2018non}} & \multirow{2}*{NCC} & \multirow{2}*{Deformable} & \multirow{2}*{MR} & \multirow{2}*{Brain} & 8-layer \\
			&  &  &  &  & FCN \\ \hline
             
			\cite{cao2018deep} & NCC & Deformable & CT, MR & Pelvis & U-Net \\
			\hline
			
			\multirow{2}*{\cite{de2017end}} & \multirow{2}*{NCC} & \multirow{2}*{Deformable} & \multirow{2}*{MR} & Cardiac & \multirow{2}*{DIRNet} \\
			&  &  &  & Cine & \\ \hline
			
			\multirow{2}*{\cite{de2018deep}} & \multirow{2}*{NCC} & \multirow{2}*{Deformable} & \multirow{2}*{MR} & Cardiac & \multirow{2}*{DLIR} \\
			&  &  &  & Cine & \\ \hline
			
			\multirow{3}*{\cite{ferrante2018adaptability}} & \multirow{3}*{NCC} & \multirow{3}*{Deformable} & \multirow{3}*{X-ray, MR} & Bone & \multirow{2}*{U-Net} \\
			&  &  &  & Cardiac & \multirow{2}*{STN} \\ 
			&  &  &  & Cine &  \\
			\hline
			
			\multirow{2}*{\cite{sun2018deformable}} & L2 Distance + & \multirow{2}*{Deformable} & \multirow{2}*{MR, US} & \multirow{2}*{Brain} & \multirow{2}*{FCN} \\
			& Image Gradient &  &  &  &  \\ \hline
			
            \cite{neylon2017neural} & Predicted TRE & Deformable & CT & Head/Neck & FCN \\
			\hline
            
			\cite{fan2018adversarial} & BCE & Deformable & MR & Brain & GAN \\
            \hline

			\multirow{2}*{\cite{mahapatra2018elastic}} & NMI + SSIM & \multirow{2}*{Deformable} & MR, FA/ & Cardiac & \multirow{2}*{GAN} \\
			& + VGG Outputs &  & Color fundus & Retinal & \\ \hline
			  
             \multirow{3}*{\cite{mahapatra2018joint}} & NMI + SSIM +& \multirow{3}*{Deformable} & \multirow{3}*{X-ray} & \multirow{3}*{Bone} & \multirow{3}*{GAN} \\
			  & VGG Outputs +&  &  &  & \\
			  & BCE &  &  &  & \\\hline
 
 			 \multirow{2}*{\cite{yoo2017ssemnet}} & \multirow{2}*{MSE AE Output} & \multirow{2}*{Deformable} & \multirow{2}*{ssEM} & \multirow{2}*{Neurons} & CAE \\
			&  &  &  &  & STN \\ \hline
			
			\multirow{2}*{\cite{wu2016scalable}} & MSE Stacked & \multirow{2}*{Deformable} & \multirow{2}*{MR} & \multirow{2}*{Brain} & Stacked \\
			& AE Outputs &  &  &  & AE\\ \hline
			
			\multirow{2}*{\cite{wu2013unsupervised}} & NCC of & \multirow{2}*{Deformable} & \multirow{2}*{MR} & \multirow{2}*{Brain} & Stacked \\
			& ISA Outputs &  &  &  & ISA \\ \hline
            
            \multirow{2}*{\cite{krebs2018learning}} & \multirow{2}*{Log Likelihood} & \multirow{2}*{Deformable} & \multirow{2}*{MR} & \multirow{2}*{Brain} & cVAE \\
			&  &  &  &  & STN \\ \hline 
			
            \multirow{2}*{\cite{liu2017tensor}} & SSD MIND +& \multirow{2}*{Deformable} & \multirow{2}*{CT, MR} & Chest & FCN \\
			& PCANet Outputs &  &  & Brain & PCANet \\ \hline 
			
            \multirow{2}*{\cite{kori2018zero}} & SSD VGG & \multirow{2}*{Rigid} & \multirow{2}*{MR} & \multirow{2}*{Brain} & CNN \\
			& Outputs &  &  &  & MLP\\ \hline
		\end{tabular}
	\end{adjustbox} 
\end{table*}

\section{Unsupervised Transformation Estimation}

Despite the success of the methods described in the previous sections, the difficult nature of the acquisition of reliable ground truth remains a significant hindrance \cite{uzunova2017training}. This has motivated a number of different groups to explore unsupervised approaches \cite{de2017end, li2018non}. One key innovation that has been useful to these works is the spatial transformer network (STN) \cite{jaderberg2015spatial}. Several methods use an STN to perform the deformations associated with their registration applications \cite{ferrante2018adaptability, kuang2018faim}.
This section discusses unsupervised methods that utilize image similarity metrics (Section \ref{sec:metric}) and feature representations of image data (Section \ref{sec:Feature}) to train their networks. A description of notable works is given in Table \ref{tab:trans_unsupervised}. 


\subsection{Similarity Metric based Unsupervised Transformation Estimation}
\label{sec:metric}

\subsubsection{Standard Methods}
\label{sec:metric overview}

\begin{figure}[tbp]
	\centering
	\includegraphics[width=\columnwidth]{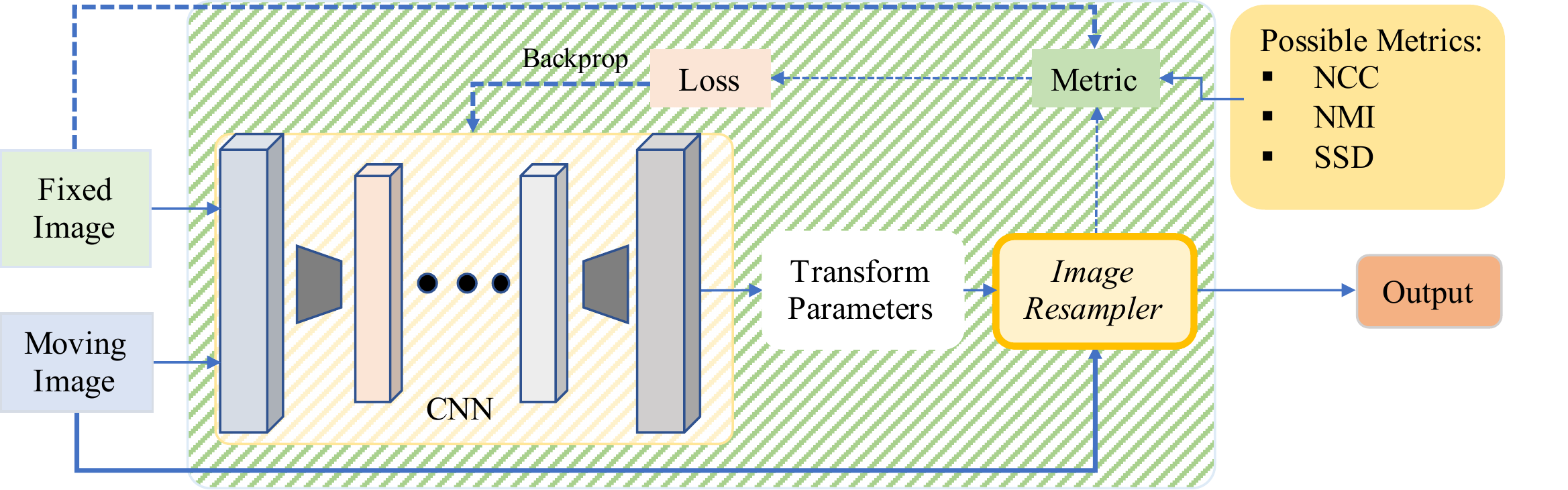}
	\caption{A visualization of deep single step registration where the network is trained using a metric that quantifies image similarity. Therefore, the approach is unsupervised.
}
	\label{fig:trans_est_unsup}       
\end{figure}

This section begins by discussing approaches that use a common similarity metric with common regularization strategies to define their loss functions. Later in the section, approaches that use more complex similarity metric based strategies are discussed. A visualization of standard similarity metric based transformation estimation is given in Fig.~\ref{fig:trans_est_unsup}.

Inspired to overcome the difficulty associated with obtaining ground truth data, Li et al. \cite{li2017non,li2018non} trained an FCN to perform deformable intersubject registration of 3D brain MR volumes using "self-supervision."
NCC between the warped and fixed images and several common regularization terms (\emph{e.g.} smoothing constraints) constitute the loss function in this method. Although many manually defined similarity metrics fail in the multimodal case (with the occasional exception of MI), they are often suitable for the unimodal case. The method detailed in this work outperforms Advanced Neuroimaging Tools (ANTs) based registration \cite{avants2011reproducible} and the deep learning methods proposed by Sokooti et al. \cite{sokooti2017nonrigid} (discussed previously) and Yoo et al. \cite{yoo2017ssemnet} (discussed in the next section).

Further, de Vos et al. \cite{de2017end} used NCC to train an FCN to perform the deformable registration of 4D cardiac cine MR volumes. A DVF is used in this method to deform the moving volume. Their method outperforms registration that is performed using the Elastix toolbox \cite{klein2010elastix}.

In another work, de Vos et al. \cite{de2018deep} use a multistage, multiscale approach to perform unimodal registration on several datasets. NCC and a bending-energy regularization term are used to train the networks that predict an affine transformation and subsequent coarse-to-fine deformations using a B-Spline transformation model. In addition to validating their multi-stage approach, they show that their method outperforms registration that is performed using the Elastix toolbox \cite{klein2010elastix} with and without bending energy.

The unsupervised deformable registration framework used by Ghosal et al. \cite{ghosal2017deep} minimizes the upper bound of the SSD (UB SSD) between the warped and fixed 3D brain MR images. The design of their network was inspired by the SKIP architecture \cite{long2015fully}. This method outperforms log-demons based registration.

Shu et al. \cite{shu2018unsupervised} used a coarse-to-fine, unsupervised deformable registration approach to register images of neurons that are acquired using a scanning electron microscope (SEM). The mean squared error (MSE) between the warped and fixed volumes is used as the loss function here. Their approach is competitive with and faster than the sift flow framework \cite{liu2011sift}.

Sheikhjafari et al. \cite{sheikhjafari2018unsupervised} used learned latent representations to perform the deformable registration of 2D cardiac cine MR volumes. Deformation fields are thus obtained by embedding. This latent representation is used as the input to a network that is composed of 8 fully connected layers to obtain the transformation. The sum of absolute errors (SAE) is used as the loss function. 
Here, the registration performance was seen to be influenced by the B-spline grid spacing.
This method outperforms a moving mesh correspondence based method described in \cite{punithakumar2017gpu}.

Stergios et al. \cite{stergios2018linear} used a CNN to both linearly and locally register inhale-exhale pairs of lung MR volumes. Therefore, both the affine transformation and the deformation are jointly estimated. The loss function is composed of an MSE term and regularization terms. Their method outperforms several state-of-the-art methods that do not utilized ground truth data, including Demons \cite{lorenzi2013lcc}, SyN \cite{avants2008symmetric}, and a deep learning based method that uses an MSE loss term. Further, the inclusion of the regularization terms is validated by an ablation study.

The successes of deep similarity metric based unsupervised registration motivated Neylon et al. \cite{neylon2017neural} to use a neural network to learn the relationship between image similarity metric values and TRE when registering CT image volumes. This is done in order to robustly assess registration performance. The network was able to achieve subvoxel accuracy in 95\% of cases. Similarly inspired, Balakrishnan et al. \cite{balakrishnan2018unsupervised, balakrishnan2018voxelmorph} proposed a general framework for unsupervised image registration, which can be either unimodal or multimodal theoretically. The neural networks are trained using a selected, manually-defined image similarity metric (\emph{e.g.} NCC, NMI, etc.).  

In a follow-up paper, Dalca et al. \cite{dalca2018unsupervised} casted deformation prediction as variational inference. Diffeomorphic integration is combined with a transformer layer to obtain a velocity field. Squaring and rescaling layers are used to integrate the velocity field to obtain the predicted deformation. MSE is used as the similarity metric that, along with a regularization term, define the loss function. Their method outperforms ANTs based registration \cite{avants2011reproducible} and the deep learning based method described in \cite{balakrishnan2018unsupervised}.

Shortly after, Kuang et al. \cite{kuang2018faim} used a CNN and STN inspired framework to perform the deformable registration of T1-weighted brain MR volumes. The loss function is composed of a NCC term and a regularization term. This method uses Inception modules, a low capacity model, and residual connections instead of skip connections. They compare their method with VoxelMorph (the method proposed by Balakrishnan et al., described above) \cite{balakrishnan2018voxelmorph} and uTIlzReg GeoShoot \cite{vialard2012diffeomorphic} using the LBPA40 and Mindboggle 101 datasets and demonstrate superior performance with respect to both.

Building upon the progress made by the previously described metric-based approaches, Ferrante et al. \cite{ferrante2018adaptability} used a transfer learning based approach to perform unimodal registration of both X-ray and cardiac cine images. In this work, the network is trained on data from a source domain using NCC as the primary loss function term and tested in a target domain. They used a U-net like architecture \cite{ronneberger2015u} and an STN \cite{jaderberg2015spatial} to perform the feature extraction and transformation estimation respectively. They demonstrated that transfer learning using either domain as the source or the target domain produces effective results. This method outperformed registration obtained using the Elastix toolbox \cite{klein2010elastix} with parameters determined using grid search.

Although applying similarity metric based approaches to the multimodal case is difficult, Sun et al. \cite{sun2018deformable} proposed an unsupervised method for 3D MR/US brain registration that uses a 3D CNN that consists of a feature extractor and a deformation field generator. This network is trained using a similarity metric that incorporates both pixel intensity and gradient information. Further, both image intensity and gradient information are used as inputs into the CNN.

\subsubsection{Extensions}

Cao et al. \cite{cao2018deep} also applied similarity metric based training to the multimodal case. Specifically, they used intra-modality image similarity to supervise the multimodal deformable registration of 3D pelvic CT/MR volumes. The NCC between the moving image that is warped using the ground truth transformation and the moving image that is warped using the predicted transformation is used as the loss function. This work utilizes "dual" supervision (\emph{i.e.} the intra-modality supervision previously described is used for both the CT and the MR images). This is not to be confused with the dual supervision strategies described earlier.

Inspired by the limiting nature of the asymmetric transformations that typical unsupervised methods estimate, Zhang et al. \cite{zhang2018inverse} used their network Inverse-Consistent Deep Network (ICNet)-to learn the symmetric diffeomorphic transformations for each of the brain MR volumes that are aligned into the same space. Different from other works that use standard regularization strategies, this work introduces an inverse-consistent regularization term and an anti-folding regularization term to ensure that a highly weighted smoothness constraint does not result in folding. Finally, the MSD between the two images allows this network to be trained in an unsupervised manner. This method outperformed SyN based registration \cite{avants2008symmetric}, Demons based registration \cite{lorenzi2013lcc}, and several deep learning based approaches.

The next three approaches described in this section used a GAN for their applications. Unlike the GAN-based approaches described previously, these methods use neither ground truth data nor manually crafted segmentations. Mahapatra et al. \cite{mahapatra2018elastic} used a GAN to implicitly learn the density function that represents the range of plausible deformations of cardiac cine images and multimodal retinal images (retinal colour fundus images and fluorescein angiography (FA) images). In addition to NMI, structual similarity index measure (SSIM), and a feature perceptual loss term (determined by the SSD between VGG outputs), the loss function is comprised of conditional and cyclic constraints, which are based on recent advances involving the implementation of adversarial frameworks. Their approach outperforms registration that is performed using the Elastix toolbox \cite{klein2010elastix} and the method proposed by de Vos et al. \cite{de2017end}.

Further, Fan et al. \cite{fan2018adversarial} used a GAN to perform unsupervised deformable image registration of 3D brain MR volumes. Unlike most other unsupervised works that use a manually crafted similarity metric to determine the loss function and unlike the previous approach that used a GAN to ensure that the predicted deformation is realistic, this approach uses a discriminator to assess the quality of the alignment. This approach outperforms Diffeomorphic Demons and SyN registration on every dataset except for MGH10. Further, the use of the discriminator for supervision of the registration network is superior to the use of ground truth data, SSD, and CC on all datasets.

Different from the hitherto previously described works (not just the GAN based ones), Mahapatra et al. \cite{mahapatra2018joint} proposed simultaneous segmentation and registration of chest X-rays using a GAN framework. The network takes 3 inputs: reference image, floating image, and the segmentation mask of the reference image and outputs the segmentation mask of the transformed image, and the deformation field. Three discriminators are used to assess the quality of the generated outputs (deformation field, warped image, and segmentation) using cycle consistency and a dice metric. The generator is additionally trained using NMI, SSIM, and a feature perceptual loss term.

Finally, instead of predicting a deformation field given a fixed parameterization as the other methods in this section do, Jiang et al. \cite{jiang2018cnn} used a CNN to learn an optimal parameterization of an image deformation using a multi-grid B-Spline method and L1-norm regularization. They use this approach to parameterize the deformable registration of 4D CT thoracic image volumes. Here, SSD is used as the similarity metric and L-BFGS-B is used as the optimizer. The convergence rate using the parameterized deformation model obtained using the proposed method is faster than the one obtained using a traditional L1-norm regularized multi-grid parameterization.

\subsubsection{Discussion and Assessment}
\label{sec:metric discuss}

Image similarity based unsupervised image registration has received a lot of attention from the research community recently because it bypasses the need for expert labels of any kind. This means that the performance of the model will not depend on the expertise of the practitioner. Further, extensions of the original similarity metric based method that introduce more sophisticated similarity metrics (\emph{e.g.} the discriminator of a GAN) and/or regularization strategies have yielded promising results. However, it is still difficult to quantify image similarity for multimodal registration applications. As a result, the scope of unsupervised, image similarity based works is largely confined to the unimodal case. Given that multimodal registration is often needed in many clinical applications, we expect to see more papers in the near future that will tackle this challenging problem.

\subsection{Feature based Unsupervised Transformation Estimation}
\label{sec:Feature}

\begin{figure}[tbp]
	\centering
	\includegraphics[width=\columnwidth]{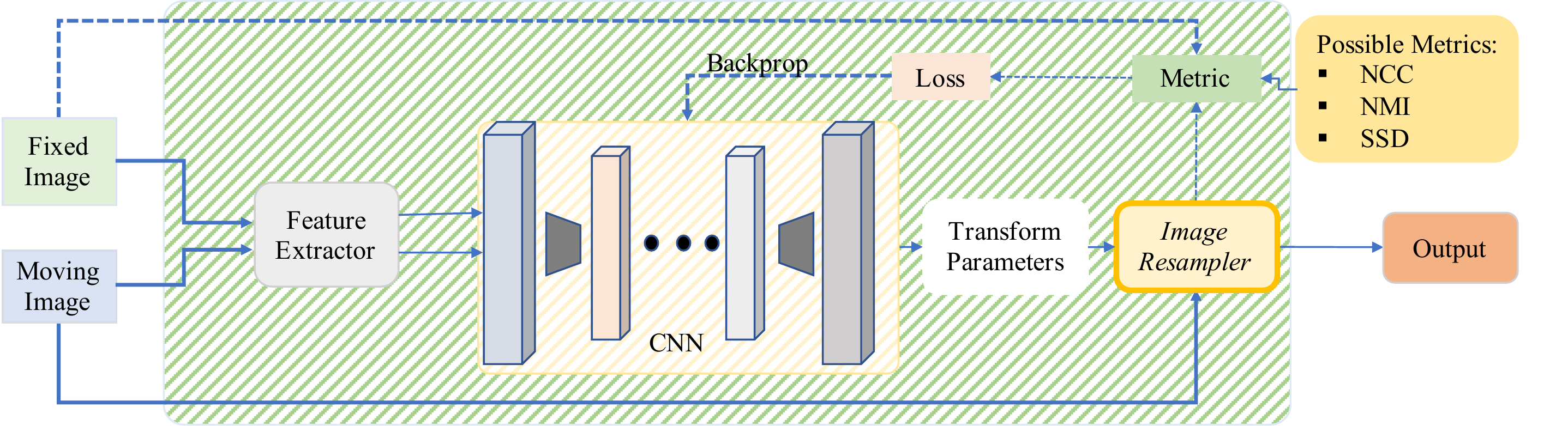}
	\caption{A visualization of feature based unsupervised image registration. Here, a feature extractor is used to map inputted images to a feature space to facilitate the prediction of transformation parameters. 
}
	\label{fig:FeatureBased}       
\end{figure}

In this section, methods that use learned feature representations to train neural networks are surveyed. Like the methods surveyed in the previous section, the methods surveyed in this section do not require ground truth data. In this section, approaches that create unimodal registration pipelines are presented first. Then, an approach that tackles multimodal image registration is discussed. A visualization of featured based transformation estimation is given in Fig.~\ref{fig:FeatureBased}.

\subsubsection{Unimodal Registration}
\label{Feature_unimodal}

Yoo et al. \cite{yoo2017ssemnet} used an STN to register serial-section electron microscopy images (ssEMs). An autoencoder is trained to reconstruct fixed images and the L2 distance between reconstructed fixed images and corresponding warped moving images is used along with several regularization terms to construct the loss function. This approach outperforms the bUnwarpJ registration technique \cite{arganda2006consistent} and the Elastic registration technique \cite{saalfeld2012elastic}.

In the same year, Liu et al. \cite{liu2017tensor} proposed a tensor based MIND method using a principle component analysis based network (PCANet) \cite{chan2015pcanet} for both unimodal and multimodal registration. Both inhale-exhale pairs of thoracic CT volumes and multimodal pairs of brain MR images are used for experimental validation of this approach. MI and residual complexity (RC) based \cite{myronenko2010intensity}, and the original MIND-based \cite{heinrich2012mind} registration techniques were outperformed by the proposed method.

Krebs et al. \cite{krebs2018learning,krebs2018unsupervised} performed the registration of 2D brain and cardiac MRs and bypassed the need for spatial regularization using a stochastic latent space learning approach. A conditional variational autoencoder \cite{doersch2016tutorial} is used to ensure that the parameter space follows a prescribed probability distribution. The negative log liklihood of the fixed image given the latent representation and the warped volume and KL divergence of the latent distribution from a prior distribution are used to define the loss function. This method outperforms the Demons technique \cite{lorenzi2013lcc} and the deep learning method described in \cite{balakrishnan2018unsupervised}.

\subsubsection{Multimodal Registration}
Unlike all of the other methods described in this section, Kori et al. perform feature extraction and affine transformation parameter regression for the multimodal registration of 2-D T1 and T2 weighted brain MRs in an unsupervised capacity using pre-trained networks \cite{kori2018zero}. The images are binarized and then the Dice score between the moving and the fixed images is used as the cost function. As the appearance difference between these two modalities is not significant, the use of these pre-trained models can be reasonably effective.

\subsubsection{Discussion and Assessment}
\label{sec:feature discuss}

Performing multimodal image registration in an unsupervised capacity is significantly more difficult than performing unimodal image registration because of the difficulty associated with using manually crafted similarity metrics to quantify the similarity between the two images, and generally using the unsupervised techniques described above to establish/detect voxel-to-voxel correspondence. The use of unsupervised learning to learn feature representations to determine an optimal transformation has generated significant interest from the research community recently. Along with the previously discussed unsupervised image registration method, we expect feature based unsupervised registration to continue to generate significant interest from the research community. Further, extension to the multimodal case (especially for applications that use image with significant appearance differences) is likely to be a prominent research focus in the next few years.

\section{Research Trends and Future Directions}
\label{sec:trends}

In this section, we summarize the current research trends and future directions of deep learning in medical image registration. As we can see from Fig.~\ref{fig:overview}, some research trends have emerged. First, deep learning based medical image registration seems to be following the observed trend for the general application of deep learning to medical image analysis. Second, unsupervised transformation estimation methods have been garnering more attention recently from the research community. Further, deep learning based methods consistently outperform traditional optimization based techniques \cite{nazib2018comparative}.
Based on the observed research trends, we speculate that the following research directions will receive more attention in the research community.

\subsection{Deep Adversarial Image Registration}

We further speculate that GANs will be used more frequently in deep learning based image registration in the next few years. As described above, GANs can serve several different purposes in deep learning based medical image registration: using a discriminator as a learned similarity metric, ensuring that predicted transformations are realistic, and using a GAN to perform image translation to transform a multimodal registration problem into a unimodal registration problem.

GAN-like frameworks have been used in several works to directly train transformation predicting neural networks. Several recent works \cite{fan2018adversarial, yan2018adversarial} use a discriminator to discern between aligned and misaligned image pairs. Although the training paradigm borrows from an unsupervised training strategy, the discriminator requires pre-aligned image pairs. Therefore, it will have limited success in multimodal or challenging unimodal applications where it is difficult to register images. Because discriminators are trained to assign all misaligned image pairs the same label, they will likely be unable to model a spectrum of misalignments. Despite this limitation, the application of GANs to medical image registration are still quite promising and will be described below.

Unconstrained deformation field prediction can result in warped moving images with unrealistic organ appearances. A common approach is to add the L2 norm of the predicted deformation field, its gradient, or its Laplacian to the loss function. However, the use of such regularization terms may limit the magnitude of the deformations that neural networks are able to predict. Therefore, Hu et al. \cite{hu2018adversarial} explored the use of a GAN-like framework to produce realistic deformations. Constraining the deformation prediction using a discriminator results in superior performance relative to the use of L2 norm regularization in that work.

Lastly, GANs can be used to map medical images in a source domain (\emph{e.g.} MR) to a target domain (\emph{e.g.} CT) \cite{choi2018stargan, isola2017image,  liu2017unsupervised, yi2017dualgan}, regardless of whether or not paired training data is available \cite{zhu2017unpaired}. This image appearance reduction technique would be advantageous because many unimodal unsupervised registration methods use similarity metrics that often fail in the multimodal case. If image translation is performed as a pre-processing step, then commonly used similarity metrics could be used to define the loss function of transformation predicting networks.

\subsection{Reinforcement Learning based Registration}

We also project that reinforcement learning will also be more commonly used for medical image registration in the next few years because it is very intuitive and can mimic the manner in which physicians perform registration. It should be noted that there are some unique challenges associated with deep learning based medical image registration: including the dimensionality of the action space in the deformable registration case. However, we believe that such limitations are surmountable because there is already one proposed method that uses reinforcement learning based registration with a deformable transformation model \cite{krebs2017robust}.

\subsection{Raw Imaging Domain Registration}

This article has focused on surveying methods performing registration using reconstructed images. However, we speculate that it is possible to incorporate reconstruction into an end-to-end deep learning based registration pipeline. In 2016, Wang~\cite{wang2016perspective} postulated that deep neural networks could be used to perform image reconstruction. Further, several works \cite{rivenson2018phase, Smith24019, yao2018deep, zhu2018image} recently demonstrated the ability of deep learning to map data points in the raw data domain to the reconstructed image domain. Therefore, it is reasonable to expect that registration pipelines that take raw data as input and output registered, reconstructed images can be developed within the next few years.

\section{Conclusion}

In this article, the recent works that use deep learning to perform medical image registration have been examined. As each application has its own unique challenges, the creation of the deep learning based frameworks must be carefully designed. Many deep learning based medical image registration applications share similar challenges including the lack of a robust similarity metric for multimodal applications, in which there are significant image appearance differences and/or different fields of view (\emph{e.g.} MR-TRUS registration) \cite{haskins2018learning}, the lack of availability of large datasets, the challenge associated with obtaining segmentations and ground truth registrations, and quantifying the uncertainty of a model's prediction. Application-specific similarity metrics, patch-wise frameworks, unsupervised approaches, and variational autoencoder inspired registration frameworks are examples of popular solutions to these challenges. Furthermore, despite the sophistication of many of the methods discussed in this survey, resampling and interpolation are often not among the components of registration that are learned by the neural network. While researchers started to pay attention to this aspect \cite{ali2019conv2warp}, we expect more works to incorporate these components into their deep learning based methods as the field continues to mature.
Recent successes have demonstrated the impact of the application of deep learning to medical image registration. This trend can be observed across medical imaging applications. Many future exciting works are sure to build on the recent progress that has been outlined in this paper.





\bibliographystyle{apa}      
\bibliography{refs}   

\end{document}